# ON THE RELEVANCE OF BANDWIDTH EXTENSION FOR SPEAKER IDENTIFICATION[1]

*Marcos Faúndez-Zanuy, Mattias Nilsson (\*), W. Bastiaan Kleijn (\*)*

Escola Universitària Politècnica de Mataró (UPC) SPAIN

(\*) Department of Speech, Music and Hearing (KTH) SWEDEN
e-mail: faundez@eupmt.es, mattiasn@speech.kth.se, and bastiaan@speech.kth.se

## ABSTRACT

In this paper we discuss the relevance of bandwidth extension for speaker identification tasks. Mainly we want to study if it is possible to recognize voices that have been bandwith extended. For this purpose, we created two different databases (microphonic and ISDN) of speech signals that were bandwidth extended from telephone bandwidth ([300, 3400] Hz) to full bandwidth ([100, 8000] Hz). We have evaluated different parameterizations, and we have found that the MELCEPST parameterization can take advantage of the bandwidth extension algorithms in several situations.

## 1. INTRODUCTION

Humans can identify more accurately known voices if they are full band signals than if they have been filtered to telephonic bandwidth. However, most of the speaker recognition research has been performed using telephonic recordings, because it offers a wide range of applications and less competence by other biometrics identification systems (the voice can be sent in a more natural way than fingerprint, face, etc., without needing a special capturing device).

On one hand, the ability of humans for speaker recognition outperforms the state of the art automatic systems [1], mainly because people often recognize talkers by their speech habits, accent, choice of vocabulary, etc. Although modern speech recognition algorithms are getting better at detecting variations in pronunciation, these cues are still harder for machines to detect than for people. On the other hand, automatic systems can outperform humans in particular scenarios. For instance, when dealing with unfamiliar voices, short test segments, large number of speakers, etc [2]. One trivial situation would be a speech signal with a masking 50Hz tone. In this situation, the person can be unable to identify the speaker and the content of the message, while it is very easy to filter out the masking tone with an automatic system. In this situation, a very easy preprocessing algorithm is able to improve the results.

In this paper, we want to evaluate the relevance of the bandwidth for automatic speech recognition, and the influence of a bandwidth extension algorithm [3]. The motivation of this study is that the state of the art bandwidth extension algorithms improve the quality of the speech signal, and increase the ability of humans to understand the content of the message and the identity of the speaker. However, no experiments have been done on the possibility to perform an automatic identification using the speech signal processed by an extension algorithm. In this paper we show that, with a MELCEPST parameterization, it is possible to identify the speaker using the bandwidth extended signal, and in several conditions there is an improvement of the identification rates compared with the results of the equivalent narrow band speech signal.

This paper is organized as follows: section 2 describes the speaker recognition algorithm and the bandwidth extension one. Section 3 deals with the database and the set of experiments. Section 4 summarizes the results, and section 5 is devoted to the main conclusions

## 2. SPEAKER RECOGNITION AND BANDWIDTH EXTENSION ALGORITHM

This section describes the speaker recognition and the bandwidth extension algorithms.

### 2.1 Speaker recognition using CM

A covariance matrix (CM) is computed for each speaker, and an Arithmetic-harmonic sphericity measure is used in order to compare matrices [4]:

$\mu(C_j C_{test}) = \log\left(tr(C_{test} C_j^{-1}) tr(C_j C_{test}^{-1})\right) - 2\log(P)$, Where $tr$ is the trace of the matrix, and the number of parameters for each speaker is $\frac{P^2+P}{2}$ (the covariance matrix is symmetric).

For the CM model, more parameters imply a higher dimensional cepstral vectors.

### 2.2 Bandwidth extension

A speech signal that has passed through the public switched telephony network (PSTN) has generally a limited frequency range between 0.3 and 3.4 kHz. This narrow-band speech signal is perceived as muffled compared to the original wide-band (0 – 8 kHz) speech signal. The Bandwidth extension algorithms aim at recovering the lost low- (0 - 0.3 kHz) and/or high- (3.4 –8 kHz) frequency band given the narrow-band speech signal. There are various techniques used for extending the bandwidth of the narrow-band. For instance, vector quantizers can be used for mapping features (e.g. parameters describing the spectral envelope) of the narrow-band to features describing the low- or high-band [9,10]. The method used in this work is based on statistical modeling between the narrow- and high-band [3]. The core of this system is a Gaussian mixture model (GMM) which models the true joint probability density function (p.d.f.) between the narrow- and high-band feature vectors. Both the narrow- and high-band feature vectors consists of a number of parameters describing the spectral envelope of respective band. In addition, the narrow-band feature vector has one parameter being a measure on the degree of voicing, and the high-band feature vector has one parameter being the difference in logarithm energy between the two frequency bands (energy-ratio).

---

[1] This work has been supported by the CICYT TIC2000-1669-C04-02

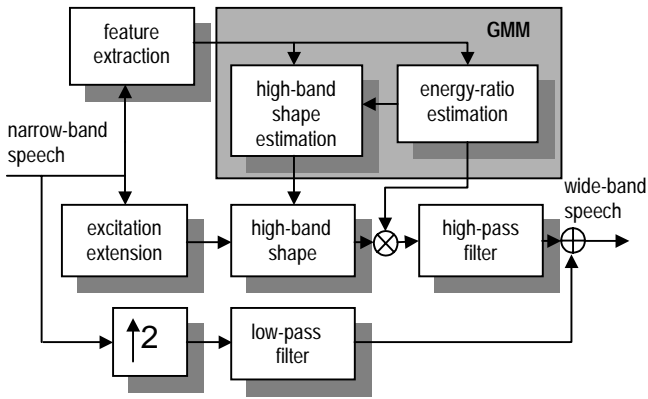

Fig 1. System architecture

A schematic of the bandwidth extension algorithm is depicted in Fig 1. From the narrow-band speech, 15 cepstral-coefficients and the degree of voicing are derived. Then, using an asymmetric (penalizing high-band energy over-estimates) cost-function together with the a-posteriori distribution of the energy-ratio given the narrow-band feature vector, we obtain an estimate of the energy-ratio between the high- and narrow-band. Given the estimated energy-ratio and the derived narrow-band features, an MMSE estimate of the spectral envelope of the high-band is computed. A modified spectral folding excitation is then filtered with the energy-ratio controlled high-band envelope and added to the up-sampled low-pass filtered narrow-band speech signal to form a reconstructed wide-band speech signal.

### 2.3 Database

Our experiments have been computed with the Gaudi database [5]. This database was previously used in [6] and [7]. Two subcorpora have been used:

a) ISDN: 43 speakers acquired with a PC connected to an ISDN. Thus, the speech signal is A law encoded at fs=8kHz, 8 bit/sample and the bandwidth is 4kHz.

b) MIC: 49 speakers acquired with a simultaneous stereo recording with two different microphones (AKG C-420 and SONY ECM66B). The speech is in wav format at fs=16kHz, 16 bit/sample, and the bandwidth is 8kHz.

The bandwidth extension algorithm has been tuned for speech signals with POTS (plain old telephone service) bandwidth, inside the range [300,3400]. For this reason, we have created the following databases (see table 1):

| Name | BW (kHz) | fs (kHz) | bps | description |
|---|---|---|---|---|
| ISDN | [0, 4] | 8 | 8 | Original |
| ISDNb | [0.3, 3.4] | 8 | 8 | ISDN filtered with potsband |
| ISDNc | [0.1, 8] | 8 | 8 | ISDNb + bandwidth extension |
| MIC | [0, 8] | 16 | 16 | Original |
| MICb | [0.3,3.4] | 16 | 16 | MIC filtered with potsband |
| MICc | [0.1, 8] | 16 | 16 | MICb + bandwidth extension |

Table 1: speech databases, where BW=bandwidth, fs=sampling frequency, bps= bits per sample

We have used the *potsband* routine, that can be downloaded in [8]. This function meets the specifications of G.151 for any sampling frequency, and has a gain of -3dB at the passband edges.

### 3. RESULTS

The speech signals are pre-emphasized by a first order filter whose transfer function is $H(z)=1-0.95z^{-1}$. A 30 ms Hamming window is used, and the overlapping between adjacent frames is 2/3. One minute of read text is used for training, and 5 sentences for testing (each sentence is about two seconds long). We have evaluated the results with two classical parameterizations:

- LPCC: Cepstral coefficients obtained with a recursion over the LPC coefficients
- MELCEPST: Cepstral Coefficients obtained with mel filters over fft spectrum.

### 3.1 Results using the ISDN database

Figure 2 shows the obtained results for several LPCC vector dimensions. We have used two different frame lengths for the bandwidth extended signal: 30ms (equivalent to 240 samples for a sampling rate of 8kHz, and 480 samples for a sampling rate of 16 kHz), and 15ms (same number of samples per frame than the original ISDN database).

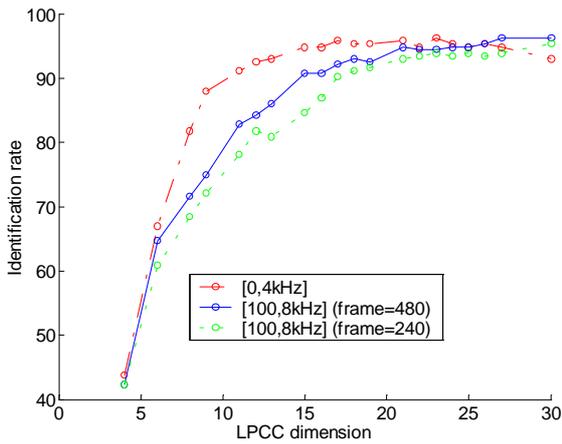

Fig. 2 Identification rates for ISDN and ISDNc as function of P, for LPCC parameterization.

It is interesting to observe that the identification rates are worse (about 10%) using the bandwidth extended signal for the usual range of the parameter vector dimension. However, we must take into account that the ISDNc database (see table 1) has been stored in 8 bit A-law format, so quantization noise has been added during this process. For this reason, it can not be concluded that the reduction on the identification rate is due to the bandwidth extension algorithm.

Figure 3 shows the obtained results for several MELCEPST vector dimensions, in similar conditions than figure 2. We have used a suitable frame length for fft computation (256 and 512 instead of 240 and 480).

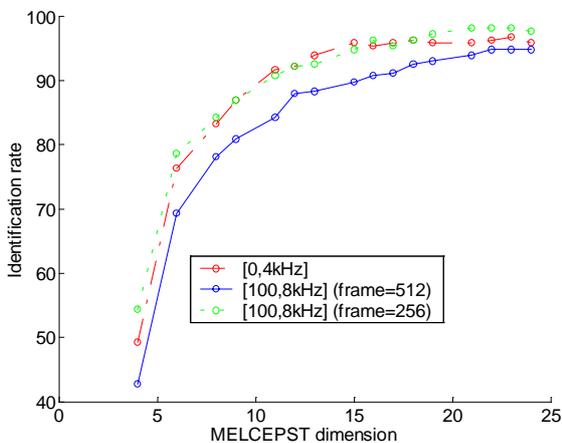

Fig. 3 Identification rates for ISDN and ISDNc as function of P, for MELCEPST parameterization.

In this case, we observe that there is a reduction in the identification rates when the frame temporal length of the bandwidth extended signal is the same than the value used for the original ISDN database. However, there is an improvement for the lowest and highest vector dimensions when the frame length is reduced. We believe that two different facts can be responsible of this behaviour:

1. The number of feature vectors is doubled when the frame length is reduced, yielding a better statistical characterization of the speaker. This can be important for pattern recognition tasks when dealing with complex distributions in P-dimensional spaces (it depends on the parameterization, P value, etc.)
2. The averaging due to the MEL filter reduces the effect of the noise. This is not true for the LPCC parameterization.

The main conclusion of this section is that MELCEPST with half frame length is the most suitable of the studied parameterizations. It achieves similar performance as without bandwidth extension.

### 3.2 Results using the MIC database

Figure 4 is analogous to figure 2, but in this case, we can compare the bandwidth extended signal with the original full band signal.

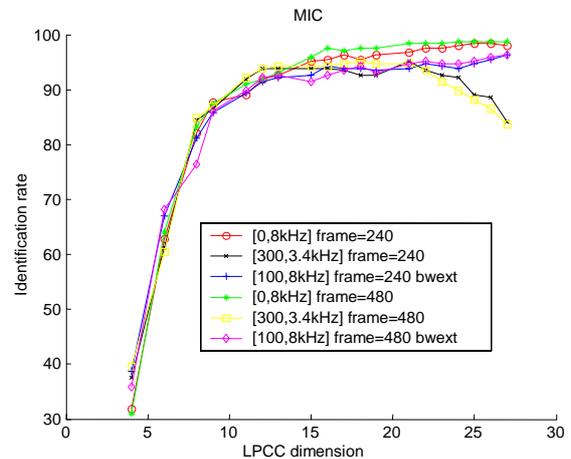

Fig. 4. Identification rates for MIC, MICb, MICc as function of P, for LPCC parameterization.

We observe that there is a significative reduction of the identification rates for the highest vector dimensions and the PSTN bandwidth. On the other hand, the bandwidth extended signal achieves similar performance to the original full band signal.

The last experiment consists on the evaluation of the MELCEPST parameterizations with the different MIC databases. Figure 5 is analogous to figure 4 but using the MELCEPST parameterization.

We observe that the bandwidth extended signal outperforms the telephonic bandwidth signal, but the identification rates are slightly smaller than the original full band signal. Nevertheless, it is a marginal improvement. Table 2 summarizes the numerical results.

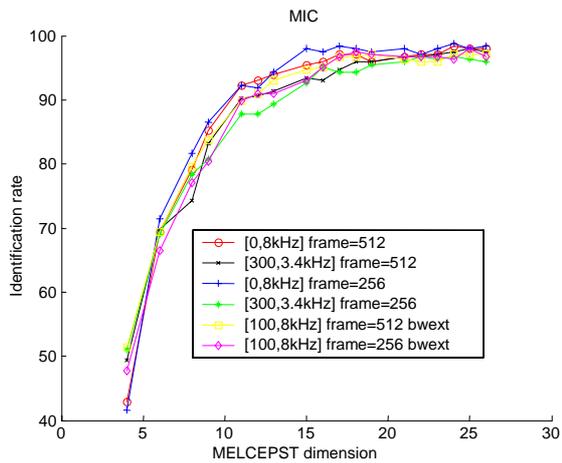

Fig. 5. Identification rates for MIC databases and MELCEPST.

|   | Frame length=512 samples | | | Frame length = 256 samples | | |
|---|---|---|---|---|---|---|
| P | MIC | MICb | MICc | MIC | MICb | MICc |
| 4 | 42.86 | 49.39 | 51.43 | 41.63 | 51.02 | 47.76 |
| 6 | 69.39 | 69.80 | 69.39 | 71.43 | 68.98 | 66.53 |
| 8 | 79.18 | 74.29 | 79.59 | 81.63 | 78.37 | 77.14 |
| 9 | 85.31 | 83.27 | 83.67 | 86.53 | 80.82 | 80.41 |
| 11 | 92.24 | 90.20 | 89.80 | 92.24 | 87.76 | 89.80 |
| 12 | 93.06 | 90.61 | 91.02 | 91.84 | 87.76 | 91.02 |
| 13 | 93.88 | 91.43 | 93.06 | 94.29 | 89.39 | 91.02 |
| 15 | 95.51 | 93.47 | 94.69 | 97.96 | 92.65 | 93.06 |
| 16 | 95.92 | 93.06 | 95.10 | 97.55 | 95.10 | 95.10 |
| 17 | 97.14 | 94.69 | 96.73 | 98.37 | 94.29 | 96.73 |
| 18 | 97.14 | 95.92 | 96.73 | 97.96 | 94.29 | 97.55 |
| 19 | 95.92 | 95.92 | 97.14 | 97.55 | 95.51 | 97.14 |
| 21 | 96.73 | 96.73 | 96.33 | 97.96 | 95.92 | 96.73 |
| 22 | 97.14 | 96.73 | 95.92 | 97.14 | 96.73 | 96.73 |
| 23 | 97.14 | 97.14 | 95.92 | 97.96 | 96.33 | 96.73 |
| 24 | 98.37 | 97.55 | 97.14 | 98.78 | 96.73 | 96.33 |
| 25 | 97.96 | 97.96 | 97.14 | 97.96 | 96.33 | 97.96 |
| 26 | 97.96 | 97.55 | 97.14 | 98.37 | 95.92 | 96.73 |

Table 2: Identification rates for MIC database and MELCEPST.

## 4. CONCLUSIONS

Although the relevance of speech coding (for instance the GSM speech encoder for mobile telephones) on a speaker recognizer has been studied by several researchers, we don't know of studies on the influence of bandwidth extension on speaker recognition. Thus, this paper is the first one devoted to the topic "what happens if the input voice to a speaker recongnizer system has been previously bandwith expanded?". A real example of a bandwith extension system is the new standard named Digital Radio Mondiale (DRM) that can be found in [11] and [12].

We have established the following conclusions:

- The studied algorithm does not introduce any damaging artifacts that affect the speaker identification rates, if a sutiable parameterization is used. Thus, we have checked the possibility to perform speaker recognition tasks using bandwith extended signals.

- The MELCEPST parameterization can take advantage of the bandwidth extension algorithm in several situations, and outperforms the LPCC. Thus, it is the recommended parameterization when a bandwidth extension algorithm is present.

- Little improvement can be achieved with a bandwidth extension algorithm in order to improve the idenditication rates of telephonic bandwith. This is quite intuitive, because no new information is added with this system. There is just a replication of the known information.